\documentclass[trackchanges, preprint2]{aastex61}

\usepackage{graphicx}
\usepackage{amsmath}
\usepackage{amsfonts}
\usepackage{amssymb}%
\providecommand{\U}[1]{\protect\rule{.1in}{.1in}}

\usepackage[yyyymmdd,hhmmss]{datetime}

\newcommand{\beq}{\begin{equation}}
\newcommand{\eeq}{\end{equation}}
\newcommand{\ba}{\begin{array}}
\newcommand{\ea}{\end{array}}

\newcommand{\ee}{\epsilon_{e,0}}

\def\be{\begin{equation}}
\def\ee{\end{equation}}

\usepackage{color}

\submitjournal{ApJL}
\shortauthors{B\'egu\'e et al.}

\begin{document}

\title{The peculiar physics of GRB 170817A and their implications for short GRBs}
\shorttitle{The Peculiar Physics of GRB 170817A}

\correspondingauthor{D. B\'egu\'e}
\email{dbegue@mpe.mpg.de}

\author[0000-0003-4477-1846]{D. B\'egu\'e}
\affiliation{Max-Planck-Institut f{\"u}r extraterrestrische Physik,
  Giessenbachstrasse, D-85748 Garching, Germany}
\author[0000-0003-3345-9515]{J. Michael Burgess}
\affiliation{Max-Planck-Institut f{\"u}r extraterrestrische Physik,
  Giessenbachstrasse, D-85748 Garching, Germany}
\affiliation{Excellence Cluster Universe, Technische Universit\"{a}t 
      M\"{u}nchen,  Boltzmannstra{\ss}e 2, 85748, Garching, Germany}
\author[0000-0003-1256-173X]{J. Greiner}
\affiliation{Max-Planck-Institut f{\"u}r extraterrestrische Physik,
  Giessenbachstrasse, D-85748 Garching, Germany}

\begin{abstract}
  The unexpected nearby gamma-ray burst GRB 170817A associated with
  the LIGO binary neutron star merger event GW170817 presents a
  challenge to the current understanding of the emission physics of
  short gamma-ray bursts (GRBs). The event's low luminosity but
  similar peak energy compared to standard short GRBs are difficult to
  explain with current models, challenging our understanding of the
  GRB emission process. Emission models invoking synchrotron radiation
  from electrons accelerated in shocks and photospheric emission are
  particularly challenging explanations for this burst.
\end{abstract}

\maketitle

\section{Introduction}

The mystery of the origin of short GRBs, the most cataclysmic events
in the Universe, has been suddenly unraveled by the nearly
simultaneous detection of gravitational waves and GRB 170817A by the
Laser Interferometer Gravitational-Wave Observatory (LIGO), by
INTEGRAL ACS and the Gamma-ray Burst monitor (GBM) on-board the Fermi
satellite \citep{LVG17}. This detection confirms the long hypothesized
scenario that some short GRBs have progenitors of binary neutron star
mergers \citep{ELP89, NPP92}. However, in the context of other known
short GRBs with measured distances, the luminosity of GRB 170817A is
$\sim$1000 times dimmer than any previously measured short GRB
\citep{GGV11}, challenging our understanding of the emission mechanism
at work.

The theoretical framework commonly invoked for GRB emission is the
fireball model \citep{Pac86,Pac90, RM92,RM94,PSN93}. It assumes that
an amount of energy comparable to the rest mass energy of the Sun is
emitted within a small region, typically few $10^{6-7}$ cm, the size
of a few solar masses black hole. The fireball is initially
optically thick, trapping the radiation within the flow. The expansion
of the fireball is triggered by its own internal thermal pressure or
by magnetic stresses, forming a jet. As the plasma expands, it becomes
transparent at a radius on the order of $10^{12}$ cm for typical GRB
parameters \citep{Goo86, Pac86, ANP91}, emitting (almost) thermal
radiation. Yet, non-thermal processes dominate the signal in the X-ray
and gamma-ray bands. In fact, it is widely believed that the spectrum
is that of synchrotron radiation emitted by electrons accelerated in
shocks or magnetic reconnection \citep{RM92, RM94, DM98,ZY11}.

The exact geometry of the relativistic outflow is subject to intense
debate, but different assumptions can be made. On one hand, its
expansion can be considered conical with little emitting material
outside of the cone opening. Those outflows are commonly referred to
as top-hat jets, as emission is only coming from the cone forming the
jet. On the other hand jets could also be structured, with a
luminosity $L$ and Lorentz factor $\Gamma$ depending only on the angle
between the center of the jet and the considered direction of
expansion \citep{RLR02, ZM02}. In this case, emission originates from
the part of the jet expanding towards the observer. Indeed,
\citet{Troja:2017}, interpret an off-axis afterglow via an observed delay 
in the X-ray
emission. Additionally, a slowly expanding cocoon can also be formed
by the material pushed away during the propagation of the jet
\citep{LDM17}.

The observed low $\gamma$-ray luminosity of GRB 170817A can be either
low intrinsic luminosity, or off-axis emission.  The first case, low
intrinsic luminosity, is very unlikely since we do not know a physical
process which can produce the same spectral shape at a luminosity
difference of 10$^5$.  In this Letter, we investigate the
  feasibility of the previously favored emission mechanisms, but just
  seen from a large off-axis angle.  We show that photospheric
emission models and synchrotron radiation produced in shocks are
particularly challenged mechanisms. After a brief description of our
analysis of GRB 170817, we consider in turn photospheric emission
models, synchrotron radiation and synchrotron self-Compton mechanisms.

\section{GBM analysis of GRB 170817A}
\begin{figure}[t]
\centering
\includegraphics[width = 0.45 \textwidth]{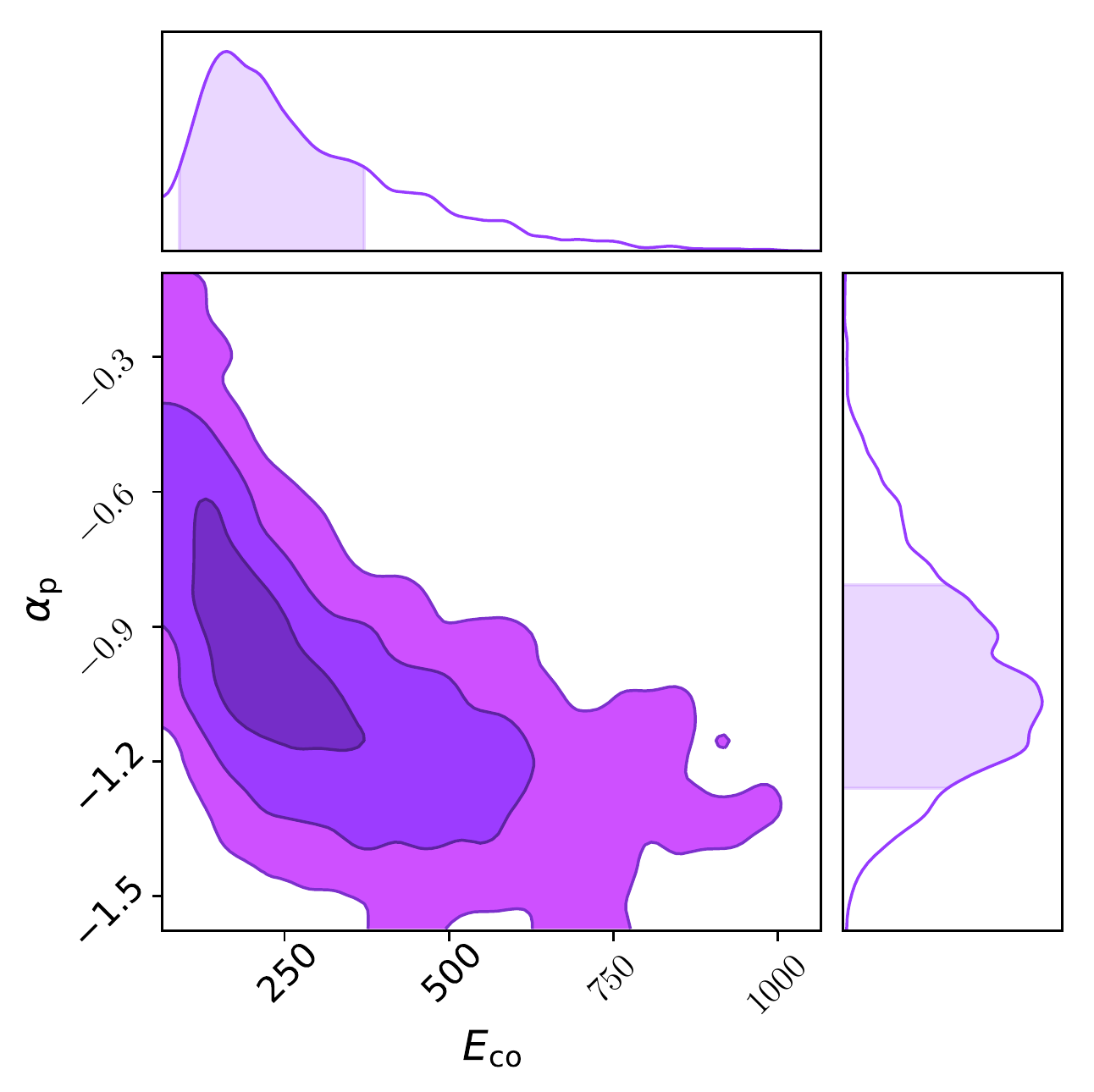}
\caption{Marginal distributions of the photon spectral index $\alpha_{\rm p}$ and of cutoff energy, $E_{\rm co} = E_{\rm p}/ (2 + \alpha_{\rm p} )$.}
\label{fig:0}
\end{figure}

Using the time-tagged Event (TTE) GBM spectral data of GRB 170817A, we
estimate the background emission from off-source regions by fitting an
unbinned Poisson-likelihood to all spectral channel count data. The
source region of the light curve is determined via a Bayesian blocks
analysis \citep{SNJ13} yielding a source region starting at T$_{0}$
-0.32 s with a dead time corrected duration of $\Delta t = 0.64$
s. With the estimated background, we perform a Bayesian spectral fit
of the source region spectrum assuming a cutoff power law model. {\tt
  MULTINEST} \citep{Feroz:2009} was used for sampling the
posterior. Informative Gaussian priors from the GBM catalogs are used
for the spectral index and peak of the spectrum resulting in a
spectral index $\alpha_{\rm p}-1 \pm 0.23$ and spectral peak
$E_p = 240_{-70}^{+130}$ keV. The posterior distributions for the peak
energy and the spectral slope are displayed in Figure \ref{fig:0}. All
analysis was performed with the Multi-Mission Maximum Likelihood
framework \citep[3ML][]{vianello:2015}.

The luminosity of the source given its distance at $\sim$40 Mpc is
9.1$\times 10^{46}$ erg s$^{-1}$. Those values are in agreement with
the analysis of \cite{LVG17}. An additional soft spectral component
was reported after the main peak. However, the statistical
significance of this emission interval 
is 1.4$\sigma$ \citep{burgess:2017}, far
below the threshold required for proper spectral analysis. As such, it
is not considered in this paper where we focus our effort in the
understanding of the physical process which produced the bright 
0.6s
long gamma-ray peak.

\section{Constraints on thermal emission}

We start our discussion by providing constraints on photospheric
emission, produced when a structured jet becomes optically thin. The
theory is simple enough such that the observed temperature and
observed flux directly constrain the Lorentz factor of the outflow and
the photospheric radius \citep{PRW07}. For GRB 170817A, we estimate
those properties to be $\Gamma \sim 65$ and
$r_{\rm ph} \sim 2.5 \times 10^8$ cm. Therefore, assuming a linear
increase of the Lorentz factor with radius, valid if the fireball is
thermally accelerated, we obtain an upper limit on the initial
expansion radius $r_0 \sim 3.5 \times 10^6$ cm. This radius is on the
order of the innermost stable circular orbit of a 3 solar mass black
hole, putting tight constraints on the plausibility of photospheric
emission. Indeed, if the jet is produced by neutrino anti-neutrino
annihilation, it requires a volume of radius several times that of the
black hole
\citep{DOB09}.
To obtain these results, we assume that the radiative efficiency at
the photosphere is 5\%. Larger radiative efficiency leads to a lower
limit on $r_0$, further reducing the possibility that the emission is
of photospheric origin. We additionally note that a similar result
holds in the case of an electron-positron pair jet, leading to a
Lorentz factor of $\Gamma \sim 5$, $r_{\rm ph} \sim 5\times 10^6$ cm
and $r_0 \sim 10^6$ cm.

\begin{figure}[t]
\centering
\includegraphics{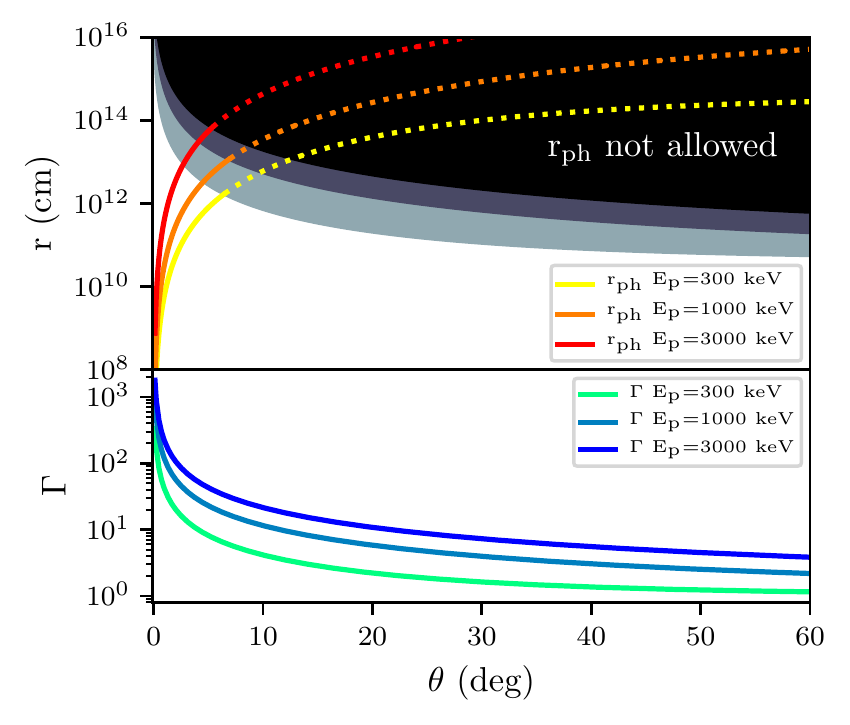}
\caption{Top: Lower-bound on the photospheric radius as a function of
  the angle for top-hat jets for several values of the on-axis (so not
  observed) spectral peak energy. The forbidden region 
  is obtained
  from requiring the curvature time to be smaller than the duration of
  GRB 170817A. Bottom: Lorentz factor of the outflow at the
  photosphere.}
\label{fig:1}
\end{figure}

We now consider that the emission is produced by a top-hat jet and
seen off axis.  Because of the relativistic motion of the outflow, the
peak energy roughly scales proportionally to the Doppler factor
$\delta = 1/(\Gamma (1-\beta \cos \theta))$, where $\beta$ is the
speed of the jet in units of the speed of light. One can show from
energy and dynamics arguments, that the observed temperature of the
photospheric component cannot be larger than the temperature close to
the black hole, which itself cannot be larger than few MeV
\citep{Goo86, Pac86}. Assuming a value for the on-axis observed peak
energy (or equivalently of the temperature), the Lorentz factor of the
outflow can be estimated\footnote{ The observed temperature is given
  by $T^{\rm obs} = \delta T^{\prime}$, while the on-axis temperature
  is given by $T^{\rm on-axis} = 2 \Gamma T^{\prime}$. Given an angle
  between the observer's direction and the direction of the jet, the
  Lorentz factor and the comoving temperature at the photosphere can
  be solved for.}.  The observed temperature translates to a lower
limit on the luminosity, which itself translates to a lower limit on
the photospheric radius. Figure \ref{fig:1} shows the value of the
Lorentz factor of the outflow and the photospheric radius $r_{\rm ph}$ as a function of
the observed angle for 3 values of the on-axis spectral peak energy:
300 keV, 1 MeV and 3 MeV. Additionally, the duration of the burst
cannot exceed the curvature time
$\Delta t_{\rm curv} = r_{\rm ph} / (2\Gamma^2 c)$, translating to an
upper limit for $r_{\rm ph}$. Here, $c$ is the speed of
light. Clearly, if the burst is seen off-axis, the angle cannot be
larger than few degrees for photospheric emission from a top-hat jet,
the least stringent limit being obtained for a peak energy of
300 keV on-axis, implying a maximum observation angle of
$\sim$8$^{\circ}$.


Another possibility is that thermal emission is produced by a nearly
spherical outflow, which expands at sub-relativistic speeds, the
so-called cocoon \citep{LDM17,KNS17}. The peak energy and the
luminosity constrain the size of the emitting region to be
$\sim 2\times 10^7$ cm. Equating the photon diffusion time to the
duration of the emission yields a number density of electrons of
$2\times 10^{27}$ cm$^{-3}$, corresponding to an associated proton
rest mass energy of a few $10^{46}$ erg s$^{-1}$. This implies that
the expansion of the cocoon is subrelativistic if pairs are not
created in a substantial amount. However, the amount of thermal energy
in the region of volume defined by the size is found to be $10^{44}$
erg, two to three orders of magnitude smaller than the total emitted
energy. This is a more contrived assumption than we wish to consider
herein. Thus, the possibility of the emission originating from a
non-relativistic cocoon is unlikely. Recently, \cite{GNP17} considered
the emission from the shock break-out of a mildly-relativistic cocoon,
and find that it explains several observables of GRB 170817A, such as
the luminosity and peak energy.

\section{Constraints on synchrotron emission}

We now proceed to constrain models for which the peak energy of the
spectrum is produced by synchrotron radiation of electrons accelerated
by shocks or magnetic reconnection. At first, we eliminate the unknown
magnetic field by using the observed spectral peak energy in the
expression for the single particle emissivity. Next, we wish to
specify the number of emitters by considering the environment in which
the jet is produced. Numerical simulations indicate that a few percent
of a solar mass ($M_{\odot}$) of material is ejected during the merger
of a neutron star binary, see \textit{e.g.} \cite{MF14}. The jet
should be energetic enough to pierce through this material and the
total energy is estimated as a fraction $\alpha$ of the solar rest
mass energy. In order to obtain a jet, $\alpha$ cannot be smaller than
the fraction of a solar mass expulsed during the merger, leading to a
value of $\alpha \sim 10^{-2 -3}$. Assuming that the outflow reaches
relativistic speeds with Lorentz factor $\Gamma$, the number of
electrons in the jet can be estimated as
$N_{tot} = E_{tot} / (\Gamma m_p c^2) = \alpha M_{\odot} / (\Gamma m_p
)$, where $E_{tot} = \alpha M_\odot c^2$ is the total
(isotropic)\footnote{For convenience, we consider only isotropic
  quantities, even knowing that the outflow is jetted. } energy
emitted, $m_p$ is the proton mass and speed of light.

Of those electrons, only a fraction $\xi$ is accelerated by
dissipative processes and emit synchrotron radiation. Provided that
pairs are not substantially created in the dissipative process, the
relation $\xi < 1$ holds. Therefore, we can write the observed
luminosity as
\begin{eqnarray}
L_{\rm obs}^{\rm sync} = \frac{4}{3} \sigma_{\rm T} c \left ( \frac{2\pi m_{\rm e} c^2}{q_{\rm e}} \right )^2 \frac{E_{\rm p}^2}{8\pi \gamma_{\rm e}^2} \frac{\alpha \xi M_\odot}{\Gamma m_{\rm p} },
\end{eqnarray}
where $\sigma_{\rm T}$ is the Thompson cross-section, $m_e$ the
electron mass, $q_{\rm e}$ is the charge of an electron and
$\gamma_{\rm e}$ is the typical electron energy.

\begin{figure}
\centering
\includegraphics[width = 0.45 \textwidth]{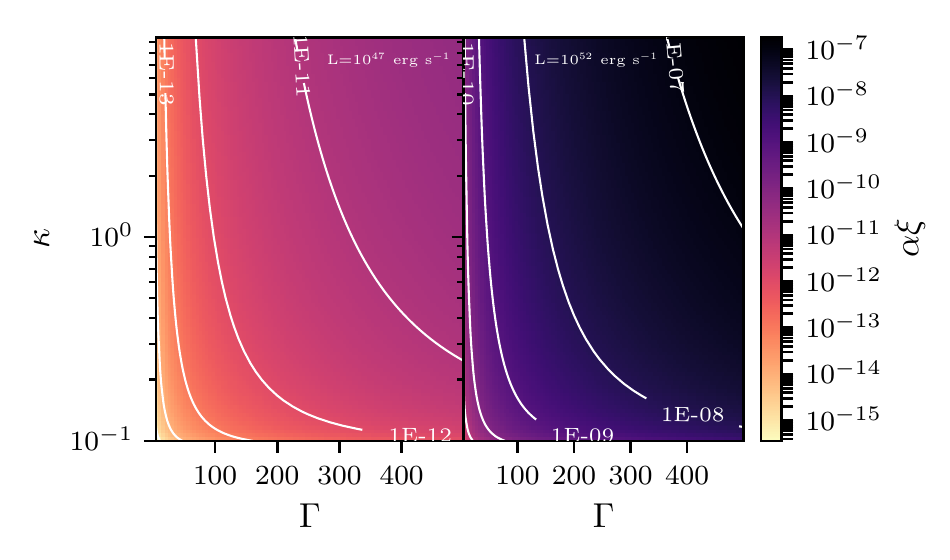}
\caption{(left) Fraction $\alpha \xi$ required by the luminosity and the peak energy for GRB 170817A as a function of $\Gamma$ and $\kappa$. (right) Same but for a GRB of luminosity $10^{52}$ erg s$^{-1}$ and a peak energy of 200 keV.}
\label{fig:2}
\end{figure}

In shock acceleration physics, the Lorentz factor of accelerated
electrons is estimated to be the ratio of the proton and electron
masses $\gamma_e = \kappa m_{\rm p} / m_{\rm e}$, where $\kappa$
parameterizes the uncertainty of the acceleration mechanism. We can
now estimate the product $\alpha \xi$ from the observed luminosity of
GRB 170817A. Figure \ref{fig:2} displays the value of $\alpha\xi$ as a
function of the Lorentz factor $\Gamma$ and $\kappa$ for observed
luminosity of GRB 170817A. A comparison is provided for a GRB with
luminosity $L = 10^{52}$ erg s$^{-1}$ and the same peak energy
$E_{\rm p} = 200$ keV. Note that $\alpha \xi$ has dependence
$E_{\rm p}^{-2}$, and therefore increasing the peak energy implies a
substantial decrease of $\alpha \xi$.

We find that the fraction $\alpha \xi$ must be extremely tiny for GRB
170817A, several orders of magnitude smaller than what is expected for
an average short GRB. There are two possibilities (and the combination
of both). First, the number of particles accelerated in the
dissipation event can be extremely small $\xi \ll 1$, leading to a
small efficiency of the radiation process. Second, the total energy in
the jet is small $\alpha \ll 10^{-2}$, which is difficult to obtain
simultaneously with a relativistic jet. Weaker constraints on
$\alpha \xi$ are obtained by increasing the Lorentz factor of the
outflow, but this is difficult because of the high baryon pollution
expected in the surrounding of the black hole immediately after the
merger. Based on these arguments, we conclude that the GRB emission is
unlikely to be synchrotron radiation from a structured jet or an
on-axis top-hat jet.

We now turn to the possibility that the emission is synchrotron
radiation from a top-hat jet, seen off-axis.
Yet, the observed high peak energy would imply an inferred
  on-axis peak energy far too large even for a typical short
  GRB. Similarly to what we demonstrated for the photospheric case,
we can assume an on-axis peak energy and compute the Lorentz factor as
function of off-axis angle.  For a 2 MeV on-axis peak (among
the largest peak energies seen in a short GRB), we find that the
Lorentz factor must drop below 20 around 10$^{\circ}$. The compactness
argument \citep{Pir99} requires the Lorentz factor of the outflow to
be at least in the order of few tens to avoid excessive pair-creation,
thus requiring the observed angle to be small.  Thus, the observed emission properties of GRB
  170817A (luminosity and peak energy) cannot be explained as off-axis
  synchrotron emission from a top-hat jet.

\section{Constraints on synchrotron self-Compton (SSC) 
emission}

We now assume that the spectral peak is the self-Compton component of
 synchrotron emission peaking near the optical range. The peak
frequency of inverse Compton is
$\nu_{\rm ssc} = 2 \gamma_{\rm e}^2 \nu_{\rm s}$, while its flux is
roughly estimated as
$F_{ssc} \sim \tau_{\rm c} \gamma_{\rm e}^2 F_{\rm s}$, where
$\tau_{\rm c} = \sigma_{\rm T} n l$ is an estimate of the opacity,
$\nu_{\rm s}$ is the synchrotron peak energy and $F_{\rm s}$ is the
synchrotron flux. Here, $l \sim c \Delta t \sim 2 \times 10^{10}$cm is
an estimate of the radial extension of the jet. In the expression of
the SSC flux, we neglected several terms of order unity, and assume
that the scattering is in the Klein-Nishina regime, which is valid at
least for photons at the observed peak. In fact, the largest
uncertainty comes from self-absorption of the synchrotron component,
which requires a more advanced treatment. Keeping in mind these
caveats, we proceed with our estimates.

The electron number density $n$ is estimated from the total number of
electrons $\alpha \xi M_\odot/(\Gamma m_p)$ divided by the volume of
the fireball $V \sim 4\pi r^2 l$, where $r \sim 10^{13-14}$ cm is the
typical radius of internal shocks.  As with the synchrotron process,
we eliminate the unknown magnetic field by using the observed peak
energy and arrive at the following estimate for the SSC luminosity
\begin{eqnarray}
& L_{obs}^{SSC} = \frac{1}{24} \frac{\sigma_{\rm T}^2 m_{\rm e}^2 c^3 }{q_{\rm e}^2 m_{\rm p}^2} \frac{E_{\rm p}^2}{\Gamma^3 \gamma_{\rm e}^6} \frac{ (\alpha \xi M_\odot)^2}{l R} \\
& \sim 1.5 \times 10^{51} (\alpha \xi)^2 \kappa^{-6} \left ( \frac{R}{10^{13}}\right )^{-1} \left ( \frac{\Gamma}{10^2}\right )^{-3}\left ( \frac{l}{10^{10}}\right )^{-1} \nonumber
\end{eqnarray}
where the strong dependencies on the Lorentz factor of the outflow
$\Gamma$ and on $\kappa$ (characterizing the minimum Lorentz factor of
the accelerated electrons) have to be noted. Equating this last result
to the observed luminosity of GRB 170817A leads to values of the parameter $\alpha \xi$ on the order of a few per thousands, more realistic considering the environment close to the black hole immediately after the merger. We still note the strong dependence on the Lorentz factor and on $\kappa$. 
The detection or non-detection of the
burst in the Fermi-LAT
instrument (un-operational due to nearing the South-Atlantic anomaly;
\citep{Collaboration:2017wr}) 
or AGILE-GRID (occulted by the Earth for the first 900 s; 
\citep{Verrecchia:2017} 
would have put additional constraints on the synchrotron self-Compton
scenario. Indeed, a second order SSC component should be produced and
peak around few hundreds of GeV. 
Finally, we note that we estimated
the synchrotron absorption frequency and find that for a large set of
parameter values, it is smaller than the synchrotron peak energy.
Thus, SSC is a potential emission process for GRB 170817A.

 
As for synchrotron emission, similar constraints on the peak energy
are obtained for synchrotron self-Compton if the emission is produced
by a top-hat jet, strongly limiting the Lorentz factor of the outflow.


\section{Conclusion}

We explored different emission mechanisms to explain the unusually
weak prompt emission of GRB 170817A. We have separately examined
structured and top-hat jets, providing constraints in both cases. For
structured jets, we find that photospheric emission is very
unlikely as it requires an initial expansion radius on the same order
or even smaller than the innermost stable circular orbit of a 3 solar
mass black hole, expected to have been created during the merger. For
completeness, we also provided the estimate in the case of a pure
electron-positron plasma, which led to a similar
conclusion. Constraints on the synchrotron emission process are
derived based on the assumption of high baryon pollution close to the
black hole after the merger. The low luminosity of GRB 170817A requires 
either that only an extremely tiny fraction of the particles in the jet
radiate, and thus, a very small radiative efficiency or that the jet
be extremely clean of baryons. 
Both these requirements are very difficult to fulfill. Indeed the 
first solution demands any form of dissipation to be suppressed, while 
the second one requires the environment in which the jet is created to 
be clear of baryons.

When considering top-hat jets seen off-axis, tight constraints on the
Lorentz factor are obtained for both the photospheric emission and
synchrotron radiation. Indeed, the sharp decrease in the Doppler boost
with viewing angle and an upper limit on the peak energy of GRBs
observed to-date lead to strict limits on either the Lorentz factor of
the outflow or on the viewing angle: both required to be small.
Thus, if GRB 170817A is an 'ordinary' burst as compared to
  other short GRBs with the exception of being observed off-axis, the
  observed properties of GRB 170817A exclude synchrotron radiation and
  photospheric emission in short-duration GRBs.

Only the estimates of the synchrotron self-Compton mechanism seem to
lead to reasonable values in terms of the peak energy, Lorentz factor,
and luminosity. A detailed study of this mechanism should be performed
to take into account synchrotron self-absorption. To conclude, GRB
170817A is peculiar in many aspects and these peculiarities limit the typical
emission processes within the standard fireball framework. We
  found that synchrotron self-Compton from a structured jet might
  explain the peculiar prompt phase of GRB 170817A. This is an
  alternative to the mildly-relativistic cocoon \citep{KNS17}. In any
  case, if the emission of GRB 170817A is typical for short GRBs in
  general, then our constraints imply that most previous suggestions
  concerning the on-axis emission of short GRBs need to be
  reconsidered.  Alternatively, if GRB 170817A was atypical,
  the coincidence with the LVC trigger is amazing, and suggests the
  discovery of a new short GRB sub-class.

\acknowledgments We thank Hans-Thomas Janka for fruitful
discussions. DB is supported by the Deutsche Forschungsgemeinschaft
(SFB 1258).

\software{3ML, pymultinest \citep{Buchner:2014}, astropy, numpy,
  Maple}

\bibliography{biblio}

\end{document}